\documentclass[runningheads]{llncs}
\usepackage{graphicx}
\begin{document}
\title{Unravelling the Topographical Organization of Brain Lesions in Variants of Alzheimer's Disease Progression}
\titlerunning{Topographical Organization of Brain Lesions in AD}

\author{Gabriel Jimenez\inst{1}\orcidID{1111-2222-3333-4444} \and 
Leopold Hebert-Stevens\inst{1}\orcidID{000} \and
Susana Boluda\inst{1}\orcidID{0000-0002-8045-2706} \and
Benoît Delatour\inst{1}\orcidID{0000-0002-9910-9932} \and
Lev Stimmer\inst{1}\orcidID{0000-0003-2800-839X} \and
Daniel Racoceanu\inst{1}\orcidID{0000-0002-9416-1803}}
\authorrunning{G. Jimenez et al.}
\institute{Sorbonne Université, Paris Brain Institute - ICM, CNRS, Inria, Inserm, AP-HP, Hôpital de la Pitié Salpêtrière, Department of Neuropathology, DMU Neuroscience, Paris, France. \email{daniel.racoceanu@sorbonne-universite.fr}}

% % Anonymous version
% \author{Anonymous\inst{1}\orcidID{abcd} \and 
% Anonymous\inst{1}\orcidID{abcd} \and
% Anonymous\inst{1}\orcidID{abcd} \and
% Anonymous\inst{1}\orcidID{abcd} \and
% Anonymous\inst{1}\orcidID{abcd} \and
% Anonymous\inst{1}\orcidID{abcd}}
% %
% \authorrunning{Anonymous et al.}
% \institute{Anonymous \email{anonymous@anonymous}}

%
\maketitle              % typeset the header of the contribution
\begin{abstract}
In this study, we proposed and evaluated a graph-based framework to assess variations in Alzheimer's disease (AD) neuropathologies, focusing on classic (\textit{cAD}) and rapid (\textit{rpAD}) progression forms. Histopathological images are converted into tau-pathology-based (i.e., amyloid plaques and tau tangles) graphs, and derived metrics are used in a machine-learning classifier. This classifier incorporates SHAP value explainability to differentiate between \textit{cAD} and \textit{rpAD}. Furthermore, we tested graph neural networks (GNNs) to extract topological embeddings from the graphs and use them in classifying the progression forms of AD. The analysis demonstrated denser networks in \textit{rpAD} and a distinctive impact on brain cortical layers: \textit{rpAD} predominantly affects middle layers, whereas \textit{cAD} influences both superficial and deep layers of the same cortical regions. These results suggest a unique neuropathological network organization for each AD variant.

\end{abstract}

\section{Introduction}
\label{sec:intro}
Alzheimer's disease (AD), a leading neurodegenerative disorder, is characterized by memory loss, cognitive decline, and behavioral changes, mainly in the elderly. The disease is believed to be caused by the accumulation of beta-amyloid proteins forming amyloid plaques and altered tau proteins leading to neurofibrillary tangles, disrupting neuronal communication. AD's complexity stems from the brain's intricate structure and the disease's heterogeneity, with patients experiencing either a rapid, aggressive pathology (\textit{rpAD}) or a slower classical decline (\textit{cAD}). Diagnosing AD, often similar to senile dementia, is typically confirmed through histopathological examination, identifying amyloid plaques and tau tangles in brain tissue, a method limited by its postmortem nature, high costs, and potential for inconsistencies in manual annotations.

% Whole Slide Images (WSI) are high-resolution scans crucial in Alzheimer's disease research as they allow the examination of the cerebral cortex's six distinct layers. Despite AI's success in disease pattern analysis, the large size of WSIs poses processing challenges, which can be addressed through innovative graph representation. A graph-based approach significantly reduces data size while preserving spatial integrity and offers an unbiased, orientation-agnostic representation. Furthermore, we can analyze tissue structure using metrics from the graphs' topological information, providing insights into the disease characteristics as reported in \cite{spie2023-jimenez}.

% % Anonymous version
Whole Slide Images (WSI) are high-resolution scans crucial in Alzheimer's disease research as they allow the examination of the cerebral cortex's six distinct layers. Despite AI's success in disease pattern analysis, the large size of WSIs poses processing challenges, which can be addressed through innovative graph representation. A graph-based approach significantly reduces data size while preserving spatial integrity and offers an unbiased, orientation-agnostic representation. Furthermore, we can analyze tissue structure using metrics from the graphs' topological information, providing insights into the disease characteristics as reported in \cite{anonymous}.

Graph Neural Networks (GNNs) have emerged as critical tools in bioinformatics and neurobiology, particularly for their effectiveness with graph-structured data, key in interpreting biological networks and neurodegenerative diseases \cite{frontiers_1,frontiers_2}. In AD research, GNNs have shown promise in patient classification and staging, with applications like the multi-class Graph Convolutional Neural Network (GCNN) for categorizing disease progression \cite{pubmed_gcnn,arxiv_adiag,pubmed_gnn,pubmed_gnn_eeg}. Their growing significance in AD detection and prognosis is also evident \cite{elsevier_graph_ad}, alongside their role in patient stratification and the advancement of individualized treatments \cite{pubmed_ai_ad,nature_ml_ad,pubmed_ai_ad}. Further reports on the performance and potential of various GNN architectures have contributed significantly to AD research \cite{arxiv_power_gnn,arxiv_pna,arxiv_comp_gnn}, with applications also in tissue abnormality characterization \cite{cvf_cgc_net,biorxiv_topo_feat,bmc_gnn_histology}, enhancing our understanding of AD's pathology.

% The primary objective of this study is to discern the underlying distinctions between rapid and classic Alzheimer's disease progression by focusing on the distribution patterns of plaques and tangles across different brain tissue layers. Moving beyond the morphological characteristics of these elements \cite{spie2023-jimenez}, we emphasize their topographical arrangement within the tissue. This approach offers novel insights into the physiopathology of AD's onset and progression, potentially leading to improved diagnostic methods and targeted therapeutic interventions (precision medicine).

% Anonymous version
The primary objective of this study is to discern the underlying distinctions between rapid and classic Alzheimer's disease progression by focusing on the distribution patterns of plaques and tangles across different brain tissue layers. Moving beyond the morphological characteristics of these elements \cite{anonymous}, we emphasize their topographical arrangement within the tissue. This approach offers novel insights into the physiopathology of AD's onset and progression, potentially leading to improved diagnostic methods and targeted therapeutic interventions (precision medicine).
\section{Dataset and Graph Construction}
\label{sec:materials}
% For this study, we used a dataset of semi-automatically annotated WSI, featuring samples from 18 patients—12 diagnosed with \textit{cAD} and 6 with \textit{rpAD}. These images were sourced from the French National Brain Biobank Neuro-CEB, with requisite consent from patients or their next of kin. Frontal lobe sections, stained using AT8 antibody to accentuate phosphorylated tau, were digitized via Hamamatsu scanners at 227 nm/pixel and 221 nm/pixel resolutions. The WSIs display a spectrum of AD pathology, varying in tau pathology severity, staining quality, and tissue preservation. The dataset also includes detailed features like the object's coordinates, area, size of plaques and tangles, and the layers to which they belong. Notably, the dataset presents a wide variance in the WSIs' and ROIs' size and the count of tangles and plaques, ranging from dozens to thousands. Despite its small size, the dataset's high quality and expert annotations form a solid base for our analysis. Figure \ref{fig:layers anotated} illustrates sample annotations for ROI, layers, and objects (i.e., plaques and tangles).

\begin{figure}[htb]
    \centering
    \includegraphics[width=0.8\textwidth]{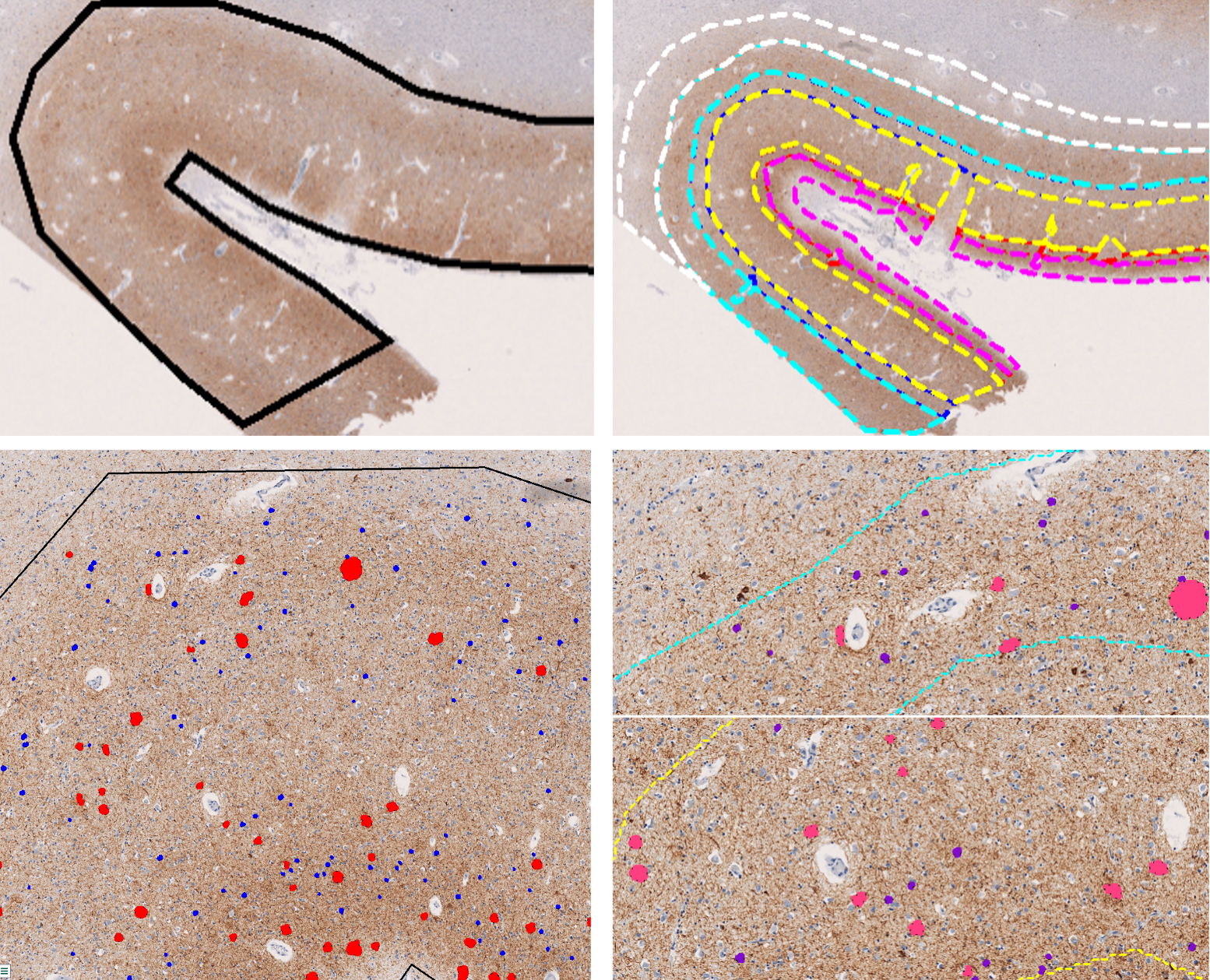}
    \caption{Example of a WSI and its annotations on the Region of Interest (ROI) on the left and the 6 layers delimitation on the right. Plaques are shown in a red mask, and tangles in a blue mask.}
    \label{fig:layers anotated}
\end{figure}

% Anonymized version
For this study, we used a dataset of semi-automatically annotated WSI, featuring samples from 18 patients—12 diagnosed with \textit{cAD} and 6 with \textit{rpAD}. These images were sourced from the Anonymous Institution, with requisite consent from patients or their next of kin. Frontal lobe sections, stained using AT8 antibody to accentuate phosphorylated tau, were digitized via Hamamatsu scanners at 227 nm/pixel and 221 nm/pixel resolutions. The WSIs display a spectrum of AD pathology, varying in tau pathology severity, staining quality, and tissue preservation. The dataset also includes detailed features like the object's coordinates, area, size of plaques and tangles, and the layers to which they belong. Notably, the dataset presents a wide variance in the WSIs' and ROIs' size and the count of tangles and plaques, ranging from dozens to thousands. Despite its small size, the dataset's high quality and expert annotations form a solid base for our analysis. Figure \ref{fig:layers anotated} illustrates sample annotations for ROI, layers, and objects (i.e., plaques and tangles).

The graphs generated focused on the spatial relationships of plaques and tangles, excluding morphological characteristics. We used the Delaunay triangulation approach, connecting points to their nearest neighbors, thereby minimizing hyperparameter selection bias. This method is ideal for biological studies due to its natural adaptation to point density and resilience against noise and outliers. In addition, we applied erosion to the graphs to isolate core structures and components within the graph. The erosion is guided by an optimal alpha value \cite{daniel_paper} derived from the data, controlling the level of granularity in the graph representation. This alpha value, set at half the median of all values from the Delaunay graph ($\alpha_{optimal} = \frac{1}{2}\cdot median(\alpha)$), ensured consistent erosion, emphasizing biologically pertinent structures and omitting less significant elements. 

Regarding the edges, those exceeding $1000 \mu m$ were excluded to align with the biological understanding that cellular interactions beyond this range are improbable. Also, we applied edge weighting inversely proportional to the square root of the distance between nodes, emphasizing the significance of proximity in cellular interactions. This approach in non-directed graphs reflects the bidirectional nature of these interactions, offering a more authentic representation of the biological networks under study.

Finally, to counter the imbalance between \textit{cAD} and \textit{rpAD} cases in the dataset, \textit{rpAD} cases were oversampled. This strategy maintained balance in model training without altering the intrinsic properties of the data, avoiding the pitfalls of conventional data augmentation methods.

\section{Methods}
\label{sec:Methods}

\subsection{Explainable Machine Learning}
\label{ssec:stat analysis}
Following the procedure described in the previous section, we created two sets of graphs for this study: one representing patient-level data and the other at the layer level representing each of the six layers in the brain cortex. Plaques and tangles were independently represented in these graphs. To effectively examine the underlying structural patterns at the patient level, each WSI-associated graph is segmented into distinct clusters, and multiple metrics (or graph features) are computed for each cluster. Analyzing data at this granular cluster level, instead of the entirety of the graph, allows for a more nuanced understanding of local spatial patterns and relationships. This granular analysis can potentially reveal differences in the spatial dispositions of plaques and tangles that might be overlooked when examining the entire graph.

Classical clustering methods such as KMeans, DBSCAN, and spectral clustering were considered inappropriate for our purposes. These algorithms primarily operate on point-based data and do not inherently consider the connectivity or relationships between nodes represented by edges in a graph. Instead, we use two alternative clustering strategies more appropriate for graph data.
\begin{itemize}
    \item \textbf{Connected Components Method:} By eroding the graph (i.e., removing certain edges), we could identify clusters or subgraphs based on direct connectivity. The benefits of this approach include its simplicity, visual clarity, and computational efficiency. However, its drawback lies in its basic clustering nature, which might miss more subtle relationships between nodes.
    \item \textbf{Markov Clustering:} This method simulates random walks on a graph to pinpoint densely connected regions. While it can discern intricate clusters, the underlying mechanism can be opaque, and it's often challenging to rationalize why certain nodes are clustered together.
\end{itemize}

We computed a range of metrics for the entire graph, layers subgraphs, and individual clusters. These metrics are essential for comparison and understanding the underlying structural and spatial relationships within the WSIs. Among the graph metrics, we computed the total number of nodes and edges, total length of edges, number of clusters, total area, and the ratio of clustered nodes to isolated nodes. Regarding the cluster metrics, we found the size of the cluster, number of edges, area, density, mean degree, dispersion, radius, inertia, alpha value optimal, and alpha value optimal normalized to the ROI. 

All the metrics mentioned above are normalized relative to the ROI size (except for the raw and normalized $\alpha$ values). This ensures consistent interpretation and minimizes potential biases introduced by size differences among WSIs. Additionally, all cluster metrics have a counterpart normalized with respect to the $\alpha_{optimal}$ value, offering a different perspective and ensuring the analysis captures nuances dependent on the alpha parameter's scale. Normalizing by the $\alpha_{optimal}$ values plays a significant role when comparing metrics on a graph-level representation. This normalization considers the $\alpha_{optimal}$ used to create the graph connections. For instance, as the number of nodes in a graph increases, the graph's area also increases. However, by normalizing using the $\alpha_{optimal}$ value, the correlation between the size of the graph and the area will vary.

These metrics were then used as features in a Random Forest (RF) classifier to distinguish between \textit{cAD} and \textit{rpAD}. The classifier's accuracy was enhanced by recursive feature elimination, analyzing feature importance, and applying SHAP values for in-depth data and model decision interpretation. We also meticulously addressed potential data leakage by recognizing and handling dependencies among clusters from the same graph.

\subsection{Deep Learning Methods}
\label{ssec:DL methods}
We also used deep learning to create meaningful clusters with node embeddings from Graph Neural Network (GNN) latent spaces, employing explainable AI (XAI) to discern model-relevant structures. A variational graph auto-encoder (VGAE) was deemed unsuitable due to the artificial nature of graph construction, leading to the selection of a GNN classifier.

\subsubsection{GNN Classifier:}
The focus was on the spatial arrangement of plaques and tangles within the WSI, independent of their morphological attributes. Therefore, 4 graph-theoretic features were introduced: the node degree, the clustering coefficient, the betweenness, and the closeness. These characteristics capture various topological properties and roles of nodes within the graph, allowing the GNN to use node-specific information for its operations. For training our GNNs, we used PyTorch Geometric and 5-fold cross-validation to ensure our results were reliable. Various GNN architectures were tested, including Graph Attention Network (GAT), Graph Convolution Network (GCN), Graph Sample and Aggregation (GraphSAGE), Weighted GraphSAGE, and Chebyshev Spectral (ChebNet). The models provided node embeddings or classifications with an early stopping mechanism to avoid overfitting. Optimal parameters were determined through hyperparameter tuning. We evaluate our model based on a weighted score of accuracy and loss from the training and validation sets ($score = 0.7 \cdot accuracy_{val} + 0.3 \cdot accuracy_{train} - 0.2 \cdot loss_{val} - 0.1 \cdot loss_{train}$). This approach helps address the limitations of our small dataset and reduces the chance of overfitting, ensuring that good validation performance is not achieved by chance only. The weights from the epoch with the highest score are saved.

% Figure \ref{fig:model} summarizes the models implemented.

% \begin{figure}[htb]
%   \centerline{\includegraphics[width=0.5\linewidth]{images/model.png}}
% \caption{Model architecture and different output options.}
% \label{fig:model}
% %
% \end{figure}

\subsubsection{Clustering on node embeddings:}
We used the 12-dimensional node embeddings extracted from the final convolution layer of our GNN for clustering purposes. These embeddings are complex representations of the nodes, reflecting the features and interrelations the GNN learned during its training. The precision with which each node's role within the network is captured is reflected in the model's accuracy. Clusters derived from these node embeddings offer greater significance and reliability than those formed directly from the graph data. This approach notably augments the Random Forest classifier's precision and depth of analysis, particularly in distinguishing between \textit{cAD} and \textit{rpAD} cases.

\subsubsection{GNN Explainer:}
GNNExplainer and PGExplainer are applied post-training to analyze the model's decision-making process. The GNNExplainer offered instance-level explanations identifying the nodes and edges relevant to the prediction. On the other hand, PGExplainer provided a broader probabilistic perspective by identifying relevant subgraphs within the input graph that contribute to the model's predictions. This dual analysis served as a form of cross-validation, building trust in the validity of our findings.
\section{Results and discussion}
\label{sec:results}
Our results showed that tangle clusters from patient-level graphs offer a more accurate classification between \textit{cAD} and \textit{rpAD}. We found an accuracy of $0.93\pm 0.08$ and $0.97\pm 0.06$ using Markov clustering and connected components clustering, respectively, compared to the $0.75\pm 0.32$ and $0.87\pm 0.2$ accuracy found when using plaques clusters and the same clustering methods respectively. When analyzing the feature importance, we found that only the tangle graphs had some features (i.e., the number of nodes and length of edges normalized to $\alpha_{optimal}$) with importance higher than 10\%. Further analysis of these features using SHAP values revealed a predominantly low feature value influence on the model's impact, although some inconsistencies were observed with higher feature values. 

Layer-specific results using RF provide an intriguing insight. Table \ref{tab:score layer} summarizes the accuracy obtained per layer. For plaques, the middle layers demonstrate higher accuracy than the outer layers, with the 3rd layer showing a 20\% increase in accuracy over the 1st and 6th layers. In contrast, the last layers, particularly the 4th, exhibit higher accuracy for tangles than the initial three layers. This variance in accuracy reveals that the most substantial differences lie in the middle layers of the brain cortex region.

\begin{table}[htb]
    \centering
    \small
    \caption{Cross-validation accuracy and standard deviation of the RF classification between \textit{cAD} and \textit{rpAD} classes for plaques and tangles layers.}
    \begin{tabular}{c|c|c|c|c|c|c}
        Layers & 1 & 2 & 3 & 4 & 5 & 6\\ \hline
        Plaques & $0.62\pm0.23$ & $0.72\pm0.18$ & $0.83\pm0.16$ & $0.67\pm0.16$ & $0.62\pm0.26$ & $0.61\pm0.23$\\
        Tangles & $0.57\pm0.16$ & $0.55\pm0.29$ & $0.64\pm0.25$ & $0.72\pm0.23$ & $0.65\pm0.22$ & $0.65\pm0.25$\\
    \end{tabular}
    \label{tab:score layer}
\end{table}

Both analyses, per-cluster and per-layer, offered valuable insights when discerning subtle differences between patient classes. In comparison, the classification using patient-level graphs yields poor performance with an accuracy of $0.67\pm 0.27$ for plaques and $0.73\pm 0.28$ for tangles. We attribute this decrease in performance to the small dataset, highlighting the limitations of studying patient-level graphs rather than more refined layer-level ones following the biological hypothesis of tau-protein aggregates progression on the brain.

Among the five GNN models tested, GCN, GAT, SAGE, SAGE with weighted edges, and CHEBnet, only GCN, SAGE, and CHEBnet yielded satisfactory performance. The GAT network provided poor results due to a lack of node features for attention calculations. On the other hand, our implementation of weighted edges in the SAGE network resulted in highly inconsistent performance, swinging between very high and very low scores due to model instability. Since each model computes the latent space
differently, resulting in different node embedding clustering; all three models with satisfactory performance were retained to test various embeddings and explore a broader range of potential clusters.

Analyzing the scores across multiple runs showed that plaque graphs yield better classification results compared to tangle graphs. While most runs for tangles score around $0.90$ accuracy, plaques achieve a higher average score of $0.98$. Consequently, analyzing clusters derived from the node embeddings of plaque graphs will likely yield more insightful results. In fact, using these embeddings increased the RF classifier's performance to $0.99\pm 0.04$ accuracy compared to the $0.87\pm 0.20$ using the Markov clustering approach. This significant improvement emphasizes the node embeddings' ability to capture key distinctions between \textit{cAD} and \textit{rpAD}. In addition, SHAP analysis revealed that certain features, notably the lower number of nodes and edge lengths combined with a higher mean degree (normalized by the alpha optimal value), strongly indicated \textit{rpAD}.

% \begin{figure}[htb]
% \centering
%  \includegraphics[width=0.7\linewidth]{images/shap_gnn.png}
% %  \vspace{2.0cm}
% \caption{SHAP values visualization of the RF differentiating between \textit{cAD} and \textit{rpAD} plaque clusters derived from GNN node embeddings. A positive impact pushes the model towards predicting \textit{rpAD}, and inversely, a negative impact to \textit{cAD}.}
% \label{fig:shap}
% %
% \end{figure}

\begin{figure}[htb]
\centering
\includegraphics[width=0.75\linewidth]{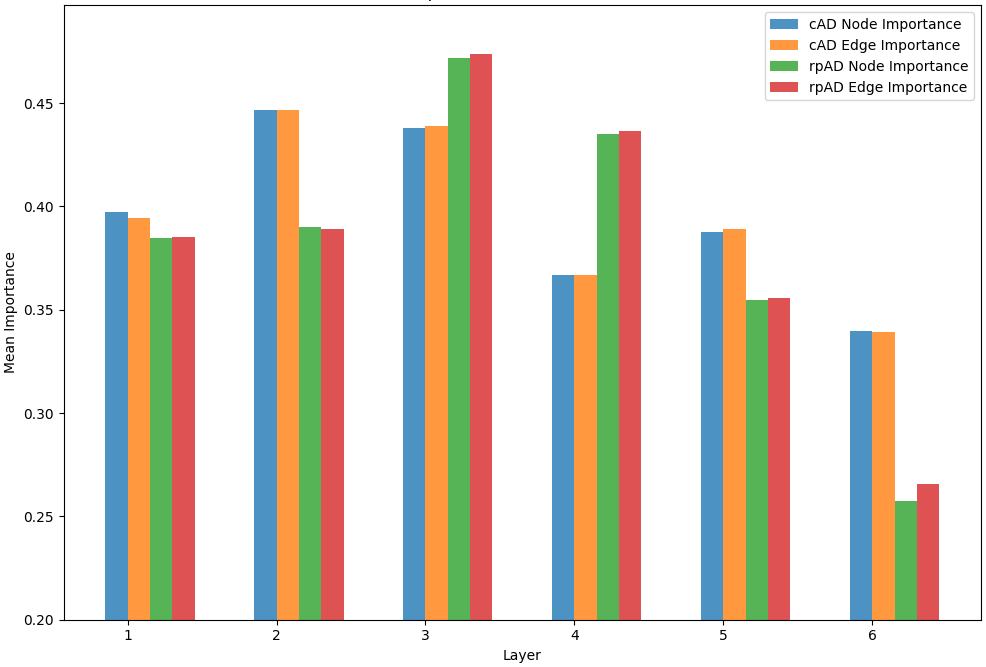}
%  \vspace{2.0cm}
\caption{Mean node and edges importance for \textit{cAD} and \textit{rpAD} for GNNExplainer.}
\label{fig:importance}
\end{figure}

The congruence in node and edge importance metrics, as identified by both GNN Explainer and PGExplainer, adds validity to our findings, suggesting they accurately represent underlying patterns. As shown in Figure \ref{fig:importance}, the importance distribution across different brain layers corroborated insights from the RF classifier, indicating that intermediate layers hold crucial discriminative information. Furthermore, the analysis revealed distinct importance levels across layers, with outer layers (2, 5, and 6) being more significant for \textit{cAD} and inner layers (3 and 4) for \textit{rpAD}. This observation implies unique structural or informational characteristics pertinent to each Alzheimer's Disease subtype at various brain depths.

\section{Conclusion}
\label{sec:discussion}
This research presents one of the first analyses of tau aggregates in layers of the brain cortex using graphs. One key observation is the denser and more interconnected graph structures in \textit{rpAD} compared to \textit{cAD}. Specifically, \textit{rpAD} exhibits compact clusters as the number of objects increases, contrasting with the expanding clusters in \textit{cAD}. Additionally, we observed distinct effects on brain layers: \textit{rpAD} predominantly affects middle layers, whereas \textit{cAD} primarily influences outer layers. These findings imply divergent cognitive impacts and clinical presentations between the two AD subtypes, mirroring each variant's unique neuropathology.

\begin{figure}[htb]
\centering
\includegraphics[width=0.48\textwidth]{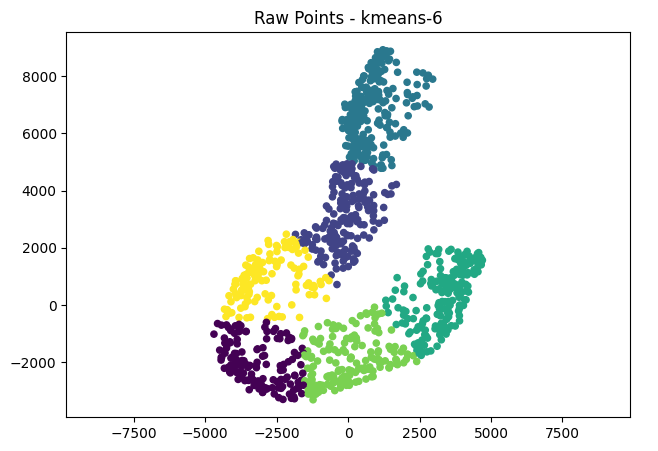}
\includegraphics[width=0.48\textwidth]{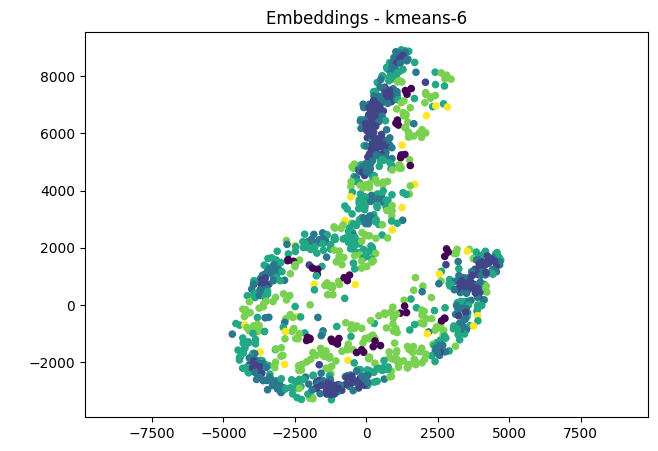}
\caption{Comparison of clustering method for raw data points and GNN-based embeddings. The top graph presents clustered nodes based on their spatial coordinates. In contrast, the bottom graph leverages GNN embeddings to cluster nodes.}
\label{fig:clustersfig}
\end{figure}

GNN methodologies proved advantageous over traditional analysis techniques, offering enriched, reliable insights into complex disease patterns. While traditional methods provide a foundational comparison, GNNs facilitate a more profound understanding of AD's intricate dynamics. For instance, GNN embeddings, as depicted in Figure \ref{fig:clustersfig}, present a more coherent spatial distribution, aligning with the brain's layered structure, in contrast to the arbitrary groupings typical of classical clustering.

% Looking forward, integrating Voronoi diagrams (complementary to Delaunay triangulation) could enhance spatial analysis. Enlarging the dataset would refine differentiation and bolster model reliability. Furthermore, incorporating three-dimensional (3D) data to reconstruct brain regions and develop 3D graphs could unveil connections obscured in two-dimensional representations, yielding more comprehensive insights. Such advancements accentuate GNNs' ability to capture spatial relationships in multi-dimensional spaces, potentially offering perspectives not fully captured by CNNs.

In conclusion, our approach combined statistical precision, machine learning's predictive capabilities, and deep learning's analytical depth to explore the nuances of AD progression. By employing a multi-faceted methodology, we aimed to ensure a thorough, wide-ranging analysis capable of discerning the subtle variances between cAD and rpAD in brain tissue. Although constrained by the dataset's size, our findings and methodologies lay the groundwork for more expansive studies, emphasizing GNNs' potential in elucidating complex neurological disorders.

\section*{Acknowledgments}
\label{sec:acknowledgments}

This research was supported by Mr Jean-Paul Baudecroux and The Big Brain Theory Program - Paris Brain Institute (ICM). The human samples were obtained from the Neuro-CEB brain bank (\url{https://www.neuroceb.org/en/}) (BRIF Number 0033-00011),  partly funded by the patients’ associations  ARSEP, ARSLA, “Connaître les Syndromes Cérébelleux,” France-DFT, France Parkinson and by Vaincre Alzheimer Foundation, to which we express our gratitude. We are also grateful to the patients and their families.

G. Jimenez and L. Hebert-Stevens contributed equally to this work. L. Stimmer is the pathologist in charge of the annotation procedure of tangles and plaques of the entire dataset used in this study.

% Anonymous version
% This research was supported by ******* and **************.

\bibliographystyle{splncs04}
\bibliography{refs}

\end{document}